\documentclass[reprint,superscriptaddress,amsmath,amssymb,aps,twocolumn,floatfix]{revtex4-1}

\usepackage[version=3]{mhchem}
\usepackage{braket}
\usepackage{dcolumn}
\newcolumntype{d}[1]{D{.}{.}{#1}}
\usepackage{graphicx}
\usepackage[normalem]{ulem}

\begin{document}

\author{Micha{\l} Hapka}
\email{michal.hapka@uw.edu.pl}
\affiliation{Faculty of Chemistry, University of Warsaw, ul.\ L.\ Pasteura 1, 02-093 Warsaw, Poland}
\author{Agnieszka Krzemi{\'n}ska}
\affiliation{Institute of Physics, Lodz University of Technology, ul.\ Wolczanska 217/221, 93-005 Lodz, Poland}
\author{Marcin Modrzejewski}
\affiliation{Faculty of Chemistry, University of Warsaw, ul.\ L.\ Pasteura 1, 02-093 Warsaw, Poland}
\author{Michał Przybytek}
\affiliation{Faculty of Chemistry, University of Warsaw, ul.\ L.\ Pasteura 1, 02-093 Warsaw, Poland}
\author{Katarzyna Pernal}
\affiliation{Institute of Physics, Lodz University of Technology, ul.\ Wolczanska 217/221, 93-005 Lodz, Poland}

\title{Efficient calculation of dispersion energy for multireference systems with Cholesky decomposition. Application to excited-state interactions}

\begin{abstract}

We propose an algorithm, that scales with the fifth power of the system size, for computing the second-order dispersion energy for monomers described with multiconfigurational wave functions. This scaling can be achieved when the number of virtual (unoccupied) orbitals largely exceeds the number of active orbitals, which is the case in practical calculations. Our approach employs Cholesky decomposition of Coulomb integrals and a recently developed recursive formula for density response functions of the monomers, enabling dispersion energy computations for systems in nondegenerate ground or excited states with arbitrary spin. As a numerical illustration, we apply the new algorithm in the framework of multiconfigurational symmetry adapted perturbation theory, SAPT(MC), to study interactions in dimers with localized excitons. The SAPT(MC) analysis reveals that the dispersion energy may be the main force stabilizing excited-state dimers.

\end{abstract}

\maketitle

\section{Introduction}\label{sec:intro}

Modelling of dispersion forces is crucial for an accurate representation of noncovalent interactions in molecular systems \cite{Grimme:11,Klimes:12} and materials \cite{French:10,Tkatchenko:15,Woods:16,Riley:10,Hobza:11}. Unfortunately, approaches to calculate the dispersion energy in excited-state complexes have been scarce \cite{Nakai:12,Sedlak:18,Ge:18a,Ge:18b,Jangrouei:22,Hancock:22}.  Semiempirical dispersion energy corrections for density functionals for ground-state complexes generally fail for dimers in excited states \cite{Besley:14,Hancock:22}. So far, two pairwise dispersion approaches have been extended to excited states. First, the local response dispersion (LRD) model of Nakai and co-workers \cite{Sato:09,Nakai:12} was applied to exciton-localized complexes from the S66 dataset \cite{Hobza:s66,Hobza:s662}. Second, Feng et al. \cite{Feng:18} used the exchange-hole dipole moment (XDM) method of Becke and Johnson \cite{Becke:05a,Becke:05b} to obtain van der Waals $C_6$ coefficients in systems involving inter- and intramolecular charge transfer excitations. The proposed generalizations of both LRD and XDM rely on excited-state electron density extracted from TD-DFT.

When ground-state interactions are concerned, accurate values of the dispersion energy can be obtained from single reference symmetry-adapted perturbation theory (SAPT) \cite{Jeziorski:94,Patkowski:20} based either on coupled-cluster \cite{Korona:08,Korona:23,Zuchowski:23} or DFT description of the monomers \cite{Misquitta:03,Hesselmann:03,Misquitta:05,Hesselmann:05}. These methods are not applicable to excited states. Recently, we have developed a wave function-based approach to the dispersion energy in ground and excited states \cite{Hapka:19a,Hapka:19b}, which employs the extended random phase approximation (ERPA) for density response \cite{Chatterjee:12}. The dispersion energy can be then predicted for any molecular system with a local exciton \cite{Jangrouei:22}. A complete description of noncovalent interactions is accessible when combining our model either with multiconfigurational SAPT \cite{Hapka:21}, SAPT(MC), or with a supermolecular approach based on multiconfigurational self-consistent field (MCSCF), in particular complete active space (CASSCF), description of the dimer \cite{Hapka:20b}. In the latter method, named CAS+DISP, the supermolecular CASSCF energy is corrected for the missing part of the dispersion energy. Both SAPT(MC) and CAS+DISP have already proven useful in studying excited-state organic dimers \cite{Jangrouei:22}. 

Currently, the bottleneck in both SAPT(MC) and CAS+DISP is the calculation of coupled dispersion and exchange-dispersion energy contributions. The computational cost of both components grows formally with the sixth power of the system size. The goal of this work is to extend the applicability of SAPT(MC) and CAS+DISP methods to larger systems by reducing the scaling of the coupled dispersion energy from $m^6$ to $m^5$. For this purpose, a novel algorithm is proposed. It employs a Cholesky decomposition technique and the recently introduced recursive formula for computation of density response functions \cite{Drwal:22}. The $m^5$ scaling is achievable if the interacting monomers are described with multiconfigurational (MC) wave functions, e.g., CASSCF, and the number of active orbitals is much smaller than that of the virtual ones, which is typically the case. The new developments are applied to study molecular interactions in excited-state organic complexes of larger size than those affordable until now for multiconfigurational dispersion methods.

The approach for multireference functions parallels previous works focused on coupled dispersion energy computations for single-reference wave functions. In particular, SAPT based on Kohn-Sham description of the monomers, SAPT(DFT) \cite{Hesselmann:05,Misquitta:05}, may employ either the density-fitting (DF) \cite{Bukowski:05} or Cholesky decomposition \cite{Sherrill:10} techniques. The algorithm of Bukowski et al. \cite{Bukowski:05}, recently improved by Xie et al. \cite{Xie:22}, is most general and applicable to computing density response of the monomers from both local and  hybrid functionals. In the case of the exchange-dispersion energy, the computational cost remains as large as $m^5$ in single-reference SAPT ($m^6$ in the multireference case) \cite{Hapka:19b} even in the DF/Cholesky formulation \cite{Podeszwa:06,Hesselmann:14,Garcia:20a,Garcia:20b,Xie:22}.

\section{Dispersion energy with multiconfigurational wave functions at the $m^5$ cost}

The spin-summed second-order dispersion formula written in terms of monomer response properties obtained within the Extended Random Phase Approximation (ERPA) reads \cite{Hapka:19a}
\begin{equation}
E^{(2)}_{\rm disp} = -16\sum_{\nu\in A,\mu\in B}\frac{\left(  \sum_{\substack{p>q\in
A\\r>s\in B}}\left[  \mathbf{\tilde{Y}}_{\nu}^{A}\right]_{pq} \left[ \mathbf{\tilde{Y}}_{\mu}^{B}\right]_{rs}\ g_{pqrs}\right)^{2}}{\omega_{\nu}^{A} +\omega_{\mu}^{B}} ,
\label{edisp0}
\end{equation}
where $pqrs$ are natural orbitals (NOs) of the monomers. Modified two-electron integrals in the NO representation $\left\{g_{pqrs}\right\}$ are defined as
\begin{equation}
\forall_{\substack{p>q\in A\\r>s\in B}}\ \ \ g_{pqrs}=(n_{p}^{1/2}+n_{q}^{1/2}) (n_{r}^{1/2}+n_{s}^{1/2}) \left\langle pr|qs\right\rangle ,\label{g}
\end{equation}
where $\braket{pr|qs}$ are two-electron Coulomb integrals in the $\braket{12|12}$ convention, $\{n_p\}_{p\in X}$ denotes a set of natural occupation numbers of monomer $X$ ($X=A, B$), and it holds that $2\sum_{p\in X}n_{p}=N_{X}$, with $N_{X}$ being a number of electrons in monomer $X$. Transition energies $\left\{  \omega_{\nu}^{X}\right\}$ and transition vectors $\left\{\mathbf{\tilde{Y}}_{\nu}^{X}\right\}$ follow from the ERPA equation \cite{Chatterjee:12,Pernal:14}
\begin{equation}
\mathbf{\mathcal{A}}_{+}^{X}\; \mathbf{\mathcal{A}}_{-}^{X} \;
\mathbf{\tilde{Y}}_{\nu}^{X} = (\omega_{\nu}^{X})^2 \; \mathbf{\tilde{Y}}_{\nu}^{X} ,\label{ERPA}
\end{equation}

where $[\mathcal{A}^X_{\pm}]_{pq,p'q'}=\big([\mathcal{A}^X]_{pq,p'q'} \pm [\mathcal{B}^X]_{pq,p'q'} \big)/\big[(n_p^{1/2} \pm n_q^{1/2})(n_{p'}^{1/2} \pm n_{q'}^{1/2})\big]$ are hessian matrices of the monomers (see Ref.~\citenum{Pernal:14}). It should be noted that both the formula for the dispersion energy in Eq.~\eqref{edisp0} and the ERPA eigenproblem in Eq.~\eqref{ERPA} are applicable to closed and open-shell systems with monomers in arbitrary spin states.

For multireference functions based on partitioning of orbitals into the inactive (doubly occupied), active (partially occupied) and virtual (unoccupied) subsets, denoted $s_1$, $s_2$ and $s_3$, respectively, the range of the $pq$ multi-index of $\left[ \mathcal{A}^X_{\pm} \right]_{pq,p'q'}$ matrices, under the condition that $p>q$, can be split into the following subranges
\begin{equation}
\begin{matrix}
p \in s_2 \wedge q \in s_1  ,\\
p \in s_3 \wedge q \in s_1  ,\\
p \in s_2 \wedge q \in s_2  ,\\
p \in s_3 \wedge q \in s_2  . \\
\end{matrix}
\end{equation}
The same partitioning applies also for the $p'q'$ multi-index. Thus, a straightforward implementation of ERPA\ requires steps that scale as $n_{\rm OCC}^3n_{\rm SEC}^3$, where $n_{\rm OCC}= M_{s_1} + M_{s_2}$ is the number of generalized occupied orbitals and $n_{\rm SEC} = M_{s_2} + M_{s_3}$ is the number of generalized secondary orbitals ($M_{s_i}$ denotes cardinality of the set $s_i$). Evaluation of the dispersion energy formula, Eq.~\eqref{edisp0}, shares the same scaling behavior (indices $\mu,\nu$ run over all $n_{\rm OCC}n_{\rm SEC}$ eigenvectors). Below we propose an algorithm leading to a lowered, $m^{5}$ scaling.

Using the integral identity
\begin{widetext}
\begin{equation}
\frac{1}{\omega_{\nu}^{A}+\omega_{\mu}^{B}} = \frac{2}{\pi} \int_{0}^{\infty} \frac{\omega_{\nu}^{A} \, \omega_{\mu}^{B}}{\Big( \left(  \omega_{\nu}^{A}\right)^2 + \omega^2 \Big) \Big(\left(  \omega_{\mu}^{B}\right)^2 + \omega^2\Big) } \, {\rm d}\omega\ ,\ \ \ \ \ \omega_{\nu}^{A}>0,\omega_{\mu}^{B}>0 ,
\label{intid}
\end{equation}
and introducing the frequency-dependent matrix ${\bf C}^X(\omega)$
\begin{equation}
\forall_{p>q, \, p'>q' \in X}
\ \ \ [{\bf C}^{X}(\omega)]_{pq,p'q'} = 
 2\sum_{\nu} \left[ \mathbf{\tilde{Y}}_\nu^{X} \right]_{pq} 
\left[ \mathbf{\tilde{Y}}_\nu^X\right]_{p'q'} \frac{\omega_\nu^X}{\left( \omega_\nu^X \right)^2 + \omega^2 } ,
\label{eq:Cw}
\end{equation}
leads to another representation of $E^{(2)}_{\rm disp}$
\begin{equation}
E^{(2)}_{\rm disp} = -\frac{8}{\pi}\int_{0}^{\infty} {\rm d}\omega 
\sum_{\substack{p>q,p^{\prime}>q^{\prime}\in A \\ r>s,r^{\prime}>s^{\prime}\in B}} 
[{\bf C}^{A}(\omega)]_{pq,p^{\prime}q^{\prime}} 
[{\bf C}^{B}(\omega)]_{rs,r^{\prime}s^{\prime}} 
\  g_{pqrs} g_{p^{\prime}q^{\prime}r^{\prime}s^{\prime}} .
\end{equation}
\end{widetext}
The $\mathbf{C}^{X}\mathbf{(}\omega)$ matrix is equivalent to the real part of density linear response function taken with the imaginary argument $i \omega$, see Eq.~({33}) in Ref.~\citenum{pernal2014intergeminal}, and is obtained by solving the following equation \cite{Drwal:22} 
\begin{equation}
\left[  \mathbf{\mathcal{A}}_{+}^{X}\mathbf{\mathcal{A}}_{-}^{X}+\omega^{2}
\mathbf{1}\right]  \mathbf{C}^{X}\mathbf{(}\omega)=\mathbf{\mathcal{A}}_{+}^X .
\label{CX}
\end{equation} 
The modified two-electron integrals $g_{pqrs}$ of Eq.~\eqref{g} can be represented via the decomposition
\begin{equation}
g_{pqrs}=\sum_{L=1}^{N_{\rm Chol}}D_{pq,L}D_{rs,L} ,\label{CD}
\end{equation}
where vectors $\mathbf{D}_{L}$ are obtained by Cholesky decomposition of the AO Coulomb matrix followed by transformation to natural-orbital representation and scaling by $n_{p}^{1/2}+n_{q}^{1/2}$ factors. The matrix product of $\mathbf{C}^{X}\mathbf{(}\omega)$ with $\mathbf{D}$ yields a reduced-dimension intermediate
\begin{equation}
\mathbf{\tilde{C}}^{X}\mathbf{(}\omega\mathbf{)}=\mathbf{C}^{X}
\mathbf{\mathbf{(}}\omega)\mathbf{D} \label{Ctil}
\end{equation}
which allows one to write
\begin{equation}
\begin{split}
E^{(2)}_{\rm disp} &=-\frac{8}{\pi}\int_{0}^{\infty}{\rm d} \omega\ \text{Tr}\left[ \mathbf{D}^{\text{T}} \mathbf{\tilde{C}}^{A}(\omega) \mathbf{\tilde{C}}^{B}(\omega)^{\text{T}} \, \mathbf{D} \right]  .
\label{edisp}
\end{split}
\end{equation}
Notice that the obtained formula applied to excited-state systems would miss the so-called non-Casimir-Polder terms arising from negative transitions $\omega_\nu^X<0$ for which the identity in Eq.~\eqref{intid} does not hold. In Ref.~\citenum{Jangrouei:22} we have shown how to account for such terms explicitly. Upon inspection, we found that non-Casimir-Polder terms are negligible for the studied systems and they are not discussed any further.

The key step in the proposed reduced-scaling algorithm for computing dispersion energy with a multireference description of the monomer wave function assumes partitioning of a monomer Hamiltonian $\hat{H}_{X}$ into partially-correlated effective Hamiltonian, $\hat{H}^{(0)}$, and the complementary part \cite{Hapka:19a}, $\hat{H}_{X}^{\prime}=\hat{H}_{X}-\hat{H}_{X}^{(0)}$. Then, the parametric representation of the Hamiltonian is introduced
\begin{equation}
\hat{H}_{X}^{\alpha} = \hat{H}_{X}^{(0)}+\alpha\hat{H}_{X}^{\prime} ,
\label{Halpha}
\end{equation}
where the $\hat{H}_{X}^{\prime}$ operator is multiplied by the coupling constant $\alpha\in\lbrack0,1]$. There are two underlying requirements in the Hamiltonian partitioning. The first one is that the wave function describing monomer $X$ is of zeroth-order in $\alpha$ for $\hat{H}_{X}^{\alpha}$. The other condition is that scaling of the ERPA\ equations corresponding to $\hat{H}_{X}^{\alpha}$ is lowered from $m^{6}$ to $m^{5}$ at $\alpha=0$. It has been shown that a group-product-function Hamiltonian \cite{Pernal:18a,Pastorczak:18}  $\hat{H}_{X}^{(0)}$  satisfies both conditions for the MC wave function based on an ansatz assuming partitioning of orbitals into inactive (doubly occupied), active (partially occupied), and virtual (unoccupied). Notice that for a single-reference wave function, $\hat{H}_{X}^{(0)}$ can be chosen as a noninteracting Hamiltonian, as in the M{\o}ller-Plesset (MP) perturbation theory.

After employing a partitioned Hamiltonian $\hat{H}^\alpha_X$, Eq.~\eqref{Halpha}, the resulting ERPA hessian matrices $\mathbf{\mathcal{A}}_{\pm}(\alpha)$ become linear functions of $\alpha$ (notice that from now on the index $X$ in hessian and response matrices is dropped for simplicity)
\begin{equation}
\mathbf{\mathcal{A}}_{\pm}(\alpha)=\mathbf{\mathcal{A}}_{\pm}^{(0)} + \alpha\mathbf{\mathcal{A}}_{\pm}^{(1)} .\label{Alinear}
\end{equation}
By contrast, the response matrix $\mathbf{C}(\alpha,\omega)$, see Eq.~\eqref{CX}, depends on $\alpha$ in a nonlinear fashion. 
Let us represent the projected response matrix $\mathbf{\tilde{C}}(\alpha,\omega)$, Eq.~\eqref{Ctil}, as a power series expansion in the coupling constant $\alpha$ around $\alpha=0$. Truncating the expansion at the $n$th order and setting $\alpha=1$, we obtain
\begin{equation}
\mathbf{\tilde{C}}(\omega) = \sum_{k=0}^{n}\frac{1}{k!}\mathbf{\tilde{C}}^{(k)}(\omega) ,
\label{Cexp}
\end{equation}
where $\mathbf{\tilde{C}}^{(k)}(\omega)$ follows from an efficient recursive scheme derived in Ref.~\citenum{Drwal:22}
\begin{align}
\mathbf{\tilde{C}(}\omega\mathbf{)}^{(0)} &  = \mathbf{\bar{A}}_+^{(0)}\mathbf{D} , \label{C00}\\
\mathbf{\tilde{C}(}\omega\mathbf{)}^{(1)} & = \mathbf{\bar{A}}_+^{(1)}\mathbf{D} - \mathbf{\bar{A}}^{(1)}\mathbf{\tilde{C}(}\omega
\mathbf{)}^{(0)}  \ \ , \label{C01} \\
\forall_{k\geq2}\ \ \ \mathbf{\tilde{C}(}\omega\mathbf{)}^{(k)} &
= -k\mathbf{\bar{A}}^{(1)} \mathbf{\tilde{C}(}\omega\mathbf{)}^{(k-1)} \\ 
& -k(k-1)\mathbf{\bar{A}}^{(2)} \mathbf{\tilde{C}(}\omega\mathbf{)}^{(k-2)} . \label{Cn}
\end{align}
The required matrices are given by the ERPA\ matrices
$\mathbf{\mathcal{A}}_{\pm}^{(0)}$ and $\mathbf{\mathcal{A}}_{\pm}^{(1)}$
\begin{align}
\mathbf{\bar{A}}_{+}^{(0)} &  =\Lambda(\omega)\mathbf{\mathcal{A}}_{+}^{(0)} ,\\
\mathbf{\bar{A}}_{+}^{(1)} &  =\Lambda(\omega)\mathbf{\mathcal{A}}_{+}^{(1)} ,\\
\mathbf{\bar{A}}^{(1)} &  =\Lambda(\omega)\left(  \mathbf{\mathcal{A}}%
_{+}^{(0)}\mathbf{\mathcal{A}}_{-}^{(1)}+\mathbf{\mathcal{A}}_{+}%
^{(1)}\mathbf{\mathcal{A}}_{-}^{(0)}\right)  ,\\
\mathbf{\bar{A}}^{(2)} &  =\Lambda(\omega)\mathbf{\mathcal{A}}_{+}%
^{(1)}\mathbf{\mathcal{A}}_{-}^{(1)} ,
\end{align}
with
\begin{equation}
\Lambda(\omega) =\left(  \mathbf{\mathcal{A}}_{+}^{(0)}\mathbf{\mathcal{A}}_-^{(0)} + \omega^2 \mathbf{1} \right)^{-1} .\label{Lambda}
\end{equation}
Notice that by setting $n=1$ in Eq.~\eqref{Cexp} for each monomer, the
dispersion energy obtained from Eq.~\eqref{edisp} will be equivalent to the
uncoupled approximation introduced in Ref.~\citenum{Hapka:19a}. For a sufficiently large $n$, one recovers full dispersion energy (i.e., the coupled dispersion energy) \cite{Hapka:19a} numerically equal to that following from Eqs.~\eqref{edisp0}--\eqref{ERPA}.

Since the dimensions of the hessian matrices and the matrix $\mathbf{D}$ are $m^2 \times m^2$ and $m^2 \times N_{\rm Chol}$, respectively, matrix multiplications involved in Eqs.~\eqref{C00}--\eqref{Cn} scale as $m^4N_{\rm Chol} \sim m^5$. As has been shown in Ref.~\citenum{Pastorczak:18}, the matrices $\mathbf{\mathcal{A}}_{+}^{(0)}$, $\mathbf{\mathcal{A}}_{-}^{(0)}$ are block-diagonal with the largest blocks of $M_{s_2}^2\times M_{s_2}^2$ size. Consequently, the cost of inversion of the $\Lambda(\omega)$ matrix, Eq.~\eqref{Lambda}, is negligible if the number of active orbitals, $M_{s_2}$, is much smaller than that of virtual orbitals, which is usually the case in practical calculations.

A valid concern is whether expansion of the linear response function at $\alpha=0$, Eq.~\eqref{Cexp}, leads to a convergent series. Although a definite proof cannot be given, it is reasonable to expect that if a monomer wave function leads to stable ERPA equations around $\alpha=0$, then the series converges. Our numerical tests on two datasets of small, weakly-correlated dimers \cite{Korona:13,Rezac:13} have shown no convergence problems, see Supporting Information. In most cases expansion up to the order $n=8$ was sufficient to achieve a $\mu E_h$ accuracy, amounting to the mean absolute percentage error below $0.1$\% in the dispersion energy. Convergence tests carried out on larger dimers, selected from the S66 test set, both in ground and excited states have led to the same conclusions, see Table~S1 in Supporting Information. An example of the convergence of $E^{(2)}_{\rm disp}$ computed according to the procedure given by Eqs.~\eqref{edisp}--\eqref{Lambda} with the truncation order $n$ in the range from $n=1$ to $n=10$ is presented for the benzene-cyclopentane complex in Figure~S1 in Supporting Information.

The evaluation of the $\alpha$-expanded response matrix requires access to $\mathcal{A}^{(1)}_{\pm}$ hessians together with three projected $\mathbf{\tilde{C}(}\omega\mathbf{)}^{(k)}$ matrices
needed to carry out the recursion. Due to their size ($m^2 \times m^2$ and $m^2 \times N_{\rm Chol}$, respectively), for systems approaching 100 atoms these quantities can no longer fit into memory and have to be stored on disk. In this regime, the disk storage and the number of I/O operations will become the main bottleneck of the proposed approach.

Due to employing the Cholesky decomposition, the cost of computing the $\mathbf{\tilde{C}(}\omega\mathbf{)}^{(k)}$ matrices, and ultimately the dispersion energy, scales as $m^{4}N_{Chol}\sim m^{5}$. The combination of Eq.~\eqref{edisp} with the recursive scheme for $\mathbf{\tilde{C}}^{(k)}(\omega)$ is the main contribution of this work.

The scaling of second-order induction energy computations in SAPT(MC) can be reduced to $m^4$ using $\alpha$-expansion of the response functions accompanied by the Cholesky decomposition of two-electron integrals. However, it is possible to achieve such scaling without relying on the coupling constant expansion, see Supporting Information for details. For the Hartree-Fock treatment of the monomers, such an alternative approach is identical to induction energy computations in the coupled Hartree-Fock scheme, as first proposed by Sadlej \cite{Sadlej:80}.

\section{Visualisation of the dispersion energy}\label{sec:vis}

The use of Eq.~\eqref{edisp} enables spatial visualization of the dispersion interactions. Following the work of Parrish et al. \cite{Parrish:14a,Parrish:14b} and our recent development \cite{Kowalski:22}, we introduce a spatially-local descriptor of the dispersion energy based on the MC wave function description of the monomers. By inspection, it can be checked that the dispersion energy expression given in Eq.~\eqref{edisp} can be written in terms of a two-particle matrix $\mathbf{Q}$, indices of which correspond to occupied, i.e., inactive or active, orbitals localized on different monomers
\begin{equation}
E^{(2)}_{\mathrm{disp}} = \sum_{q \in A, s\in B}Q_{qs} ,
\end{equation}
where
\begin{equation}
\forall_{\substack{q\in A\\s\in B}}\ \ \ Q_{qs} = -\frac{8}{\pi} \int_{0}^{\infty} \mathrm{d}\omega \sum_{\substack{p\in A\\r\in B}}
\sum_{L=1}^{N_{\rm Chol}} D_{pq,L}W_{pq,rs}^{AB}(\omega)D_{rs,L} ,
\end{equation}
and
\begin{equation}
\forall_{\substack{pq\in A\\rs\in B}}\ \ \ W_{pq,rs}^{AB}(\omega)=
\sum_{L=1}^{N_{\rm Chol}}
\tilde{C}_{pq,L}^{A}(\omega)\tilde{C}_{rs,L}^{B}(\omega) .
\end{equation}
Such a two-particle partition of the dispersion energy can be seen as a generalization of the partitioning scheme developed in Ref.~\citenum{Parrish:14a} that was applied with uncoupled amplitudes and single-determinant wave functions.

We propose a local dispersion density function for monomer $A$
as a charge-like density, where the density of the orbital is weighted by its
contribution to the dispersion energy
\begin{equation}
Q^{A}(\mathbf{r})=\sum_{q\in s_1^A \cup s_2^A }w_q \ \rho_q(\mathbf{r}) ,\label{QA}
\end{equation}
with weights defined as
\begin{equation}
\forall_{q \in s_1^A \cup s_2^A}\ \ \ w_q = \frac{\sum_{s\in B}Q_{qs}}{N_q} .
\end{equation}
$\rho_q(\mathbf{r})$ denotes either electron density of the active electrons if $q$ refers to an active orbital localized on A
\begin{equation}
\forall_{q\in s_2^A} \ \ \ \rho_q(\mathbf{r}) = \sum_{q' \in s_2^A}
n_{q'} \ \varphi_{q'}(\mathbf{r})^{2}\ , \ \ \ \ \ N_q = \sum_{q' \in s_2^A} n_{q'} ,
\end{equation}
(notice that the sum over active orbitals includes the orbital $q$) or an orbital density, if $q$ denotes an inactive orbital 
\begin{equation}
\forall_{q \in s_1^A }\ \ \ \rho_q(\mathbf{r}) = \varphi_q(\mathbf{r})^{2}\  , \ \ \ \ \ N_q = 1 .
\end{equation}
Analogous definition can be introduced by employing natural orbitals of the monomer $B$, leading to the dispersion density function localized on $B$, $Q^{B}(\mathbf{r})$. A function  $Q^{AB}(\mathbf{r})$, defined as an average of $Q^{A}(\mathbf{r})$ and $Q^{B}(\mathbf{r})$,
\begin{equation}
Q^{AB}(\mathbf{r}) = \frac{1}{2}\left( Q^{A}(\mathbf{r}) + Q^{B}(\mathbf{r}) \right) ,
\end{equation}
collects local contributions of the natural orbitals of both monomers to the dispersion interaction and, as it should, integrates to $E^{(2)}_{\rm disp}$
\begin{equation}
E^{(2)}_{\rm disp}=\int Q^{AB}(\mathbf{r})\ \text{d}\mathbf{r} .
\end{equation}
The additional cost of obtaining the $Q^{AB}(\mathbf{r})$ descriptor is marginal compared to the cost of dispersion energy computation, as all intermediate quantities are available from the calculation of $E^{(2)}_{\mathrm{disp}}$. Since natural orbitals are typically not localized, changing the $Q^{AB}(\mathbf{r})$ representation to local orbitals should provide a more informative visualization of dispersion forces. Our aim is, however, to investigate differential maps of $Q^{AB}(\mathbf{r})$ computed for ground and excited states of the studied systems. For this purpose, natural orbitals are adequate.

\section{Computational details}
\begin{figure*}
\centering
\includegraphics[width=0.8\textwidth]{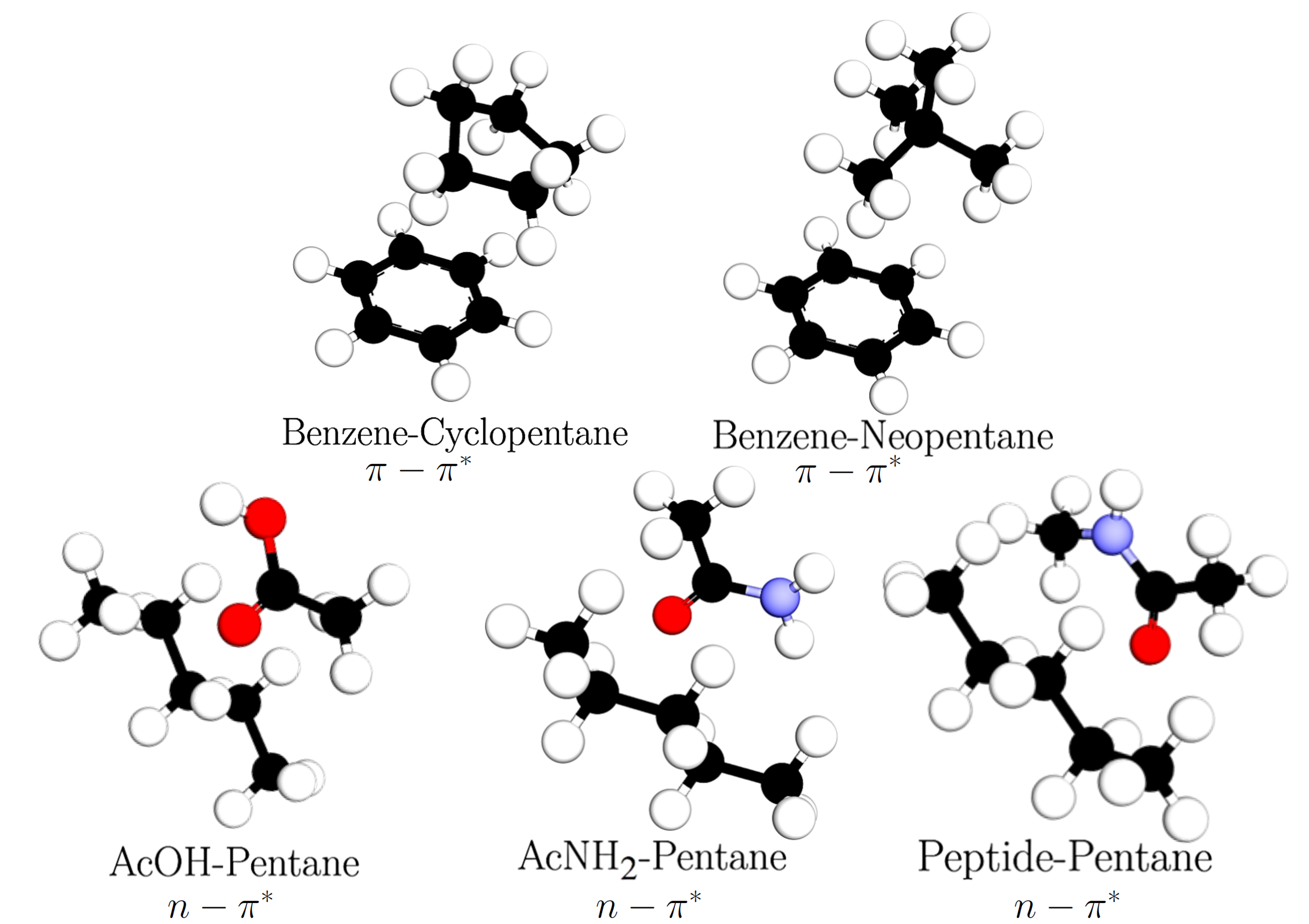}
\caption{Structures of $\pi$-$\pi^{*}$ and $n$-$\pi^{*}$ complexes analyzed in this work.}
\label{fig:geoms}
\end{figure*}

Numerical demonstration of the developed algorithm was carried out for both ground and excited states of noncovalent complexes selected from the S66 benchmark dataset of Hobza and co-workers \cite{Hobza:s66,Hobza:s662}, for which benchmark interaction energy values for electronically excited complexes have been provided by Ikabata et al. \cite{Nakai:12}.
The dimers were divided into two sets according to their size. 
Smaller systems (up to eight heavy atoms in a dimer and ca.\ 500-600 basis functions with a basis set of a triple-zeta quality) were analyzed in detail in our recent work \cite{Jangrouei:22}. Larger systems (up to eleven heavy atoms in a dimer and ca.\ 800-900 basis functions), which are beyond the capabilities of the $m^6$-implementation of the MC dispersion energy, are analyzed for the first time. This group contains five complexes: benzene-cyclopentane, benzene-neopentane, AcOH-pentane, \ce{AcNH2}-pentane, and peptide-pentane, where peptide refers to N-methylacetamide (see Figure~\ref{fig:geoms}).

Both ground- and excited-state calculations were performed using ground-state geometries taken from Ref.~\citenum{Hobza:s66}.
All supermolecular calculations employed the Boys-Bernardi counterpoise correction \cite{Boys:70}. The excitons were localized on benzene ($\pi\rightarrow\pi^{\ast}$), AcOH ($n\rightarrow\pi^{\ast}$), \ce{AcNH2} ($n\rightarrow\pi^{\ast}$), and peptide ($n\rightarrow\pi^{\ast}$) molecules. The CCSD(T) results extrapolated to the complete basis set limit (CBS) \cite{Hobza:s66} served as benchmark for ground state interaction energies. In the case of complexes involving excited states, reference results were taken from Ref.~\citenum{Nakai:12}. They were obtained by combining  CCSD(T)/CBS ground-state interaction energies with excitation energies calculated at the EOM-CCSD\cite{Monkhorst:77}/6-31++G(d,p) \cite{gaussbasis,gaussbasis2,gaussbasis3} level of theory.

The Cholesky decomposition of the Coulomb integrals matrix, $\braket{pr|qs}$, was performed in the AO basis with a modified program developed for Refs.~\citenum{Modrzejewski:20,Modrzejewski:21}. The Cholesky vectors, $R_{pq,L}$, were generated with the convergence criterion $\sum_{p \ge q} \left( \langle pp \middle| qq \rangle - \sum_L R_{pq,L} R_{pq, L} \right)<10^{-2}$, which is the same as used in Ref.~\citenum{Drwal:22}.

Second-order dispersion energies and SAPT(MC) \cite{Hapka:21} energy components based on CASSCF treatment of the monomers were computed in the \textsc{GammCor} \cite{gammcor} program. From now on SAPT(MC) based on CASSCF wave functions is denoted as SAPT(CAS). The frequency integration in Eq.~\eqref{edisp} has been carried out using 8-point Gauss-Legendre quadrature. The necessary integrals and reduced density matrices were obtained from the locally modified Molpro \cite{Molpro:12} package. Supermolecular  CASSCF and DFT-SAPT calculations were performed in Molpro. All calculations employed aug-cc-pVTZ basis set \cite{Dunning:89,Kendall:92}.

Although MP2 orbitals were used as a starting guess for the CASSCF computations, further orbitals rotations were required in almost all cases for ground- and excited-state complexes to assure that the desired orbitals are included in the active space and to maintain size-consistency in the supermolecular approach. Excited-state wave functions were computed with two-state state-averaged CASSCF. We chose the same active spaces for both ground- and excited-state calculations. The active space for benzene included  three $\pi$ bonding and the three $\pi^{*}$ antibonding MOs, which means 6 active electrons on 6 orbitals, labeled as CAS(6,6) \cite{benzene_orb}. For AcOH we chose CAS(8,8) active space including two $n$, $\pi$, $\pi^{*}$, two $\sigma$, and two $\sigma^{*}$ orbitals \cite{acoh_orb}. For \ce{AcNH2} the CAS(6,5) space was selected which involves $\sigma$, $n$, $\pi$, $\pi^{*}$, and $\sigma^{*}$ orbitals \cite{acnh2_orb}. The peptide (N-methylacetamide) active space, CAS(6,6), was composed of $\sigma$, $\pi$, $\pi^{*}$ and $\sigma^{*}$ orbitals, and two lone-pair orbitals $n$ located on oxygen atom \cite{nme_orb}.

To improve the accuracy of SAPT(CAS) interaction energy, especially for systems dominated by large polarization effects, we need to include higher than second-order induction terms. For ground-state systems which can be represented with a single Slater determinant, these terms can be approximated at the Hartree-Fock (HF) level of theory and represented as the $\delta_{\rm HF}$ correction \cite{Jeziorska:87,Moszynski:96}
\begin{equation}
\begin{split}
\delta_{\rm HF} &= E_{\rm int}^{\rm HF} - 
\big( E_{\rm elst}^{(1)}(\text{HF}) + E_{\rm exch}^{(1)}(\text{HF}) \\
&+ E_{\rm ind}^{(2)}(\text{HF}) + E_{\rm exch-ind}^{(2)}(\text{HF}) \big) ,
\end{split}
\label{deltaHF}
\end{equation}
where $E_{\rm int}^{\rm HF}$ corresponds to the supermolecular HF interaction energy and all terms are computed with Hartree-Fock wave functions. There is no straightforward way to account for higher-order polarization terms in excited-state computations. To tackle this problem, we assume that the change of higher-than-second-order induction terms upon excitation is proportional to a corresponding shift in the second-order induction, and define the $\delta_{\rm CAS}$ correction as:
\begin{equation}
\delta_{\rm CAS} = \frac{E^{(2)}_{\rm ind}({\rm ES})}{E^{(2)}_{\rm ind}({\rm GS})} \, \delta_{\rm HF} ,
\label{delta}
\end{equation}
where 
the labels GS/ES correspond to dimers in ground and excited states, respectively. Notice that in our previous work \cite{Jangrouei:22} a similar scaling expression involved sums of induction and exchange-induction ($E^{(2)}_{\rm exch-ind}$) terms. In this work, the latter is not computed directly, but follows from approximate scaling relation, see below. Such a treatment of $E^{(2)}_{\rm exch-ind}$ energy could introduce additional error in the $\delta_{\rm CAS}$ term and we decided not to include it in Eq.~\eqref{delta}.

Compared to single-reference SAPT schemes, in multiconfigurational SAPT it is not straightforward to apply the Cholesky decomposition to second-order exchange energy components, i.e., $E_{\rm exch-ind}^{(2)}$ and $E_{\rm exch-disp}^{(2)}$. The difficulty follows from the necessity to obtain separately lower and upper triangles of transition density matrices (see the discussion in Sec.~2 of Ref.~\citenum{Hapka:19b}); possible solutions will be addressed in our future work. To account for both exchange-induction and exchange-dispersion terms in this study, we propose a simple scaling scheme 
\begin{align}
E^{(2)}_{\rm exch-ind}(\rm aVTZ) = E^{(2)}_{\rm exch-ind}(\rm aVDZ) \times \frac{E^{(2)}_{\rm ind}(\rm aVTZ)}{E^{(2)}_{\rm ind}(\rm aVDZ)}  \label{exch-ind-gs} ,\\
E^{(2)}_{\rm exch-disp}(\rm aVTZ) = E^{(2)}_{\rm exch-disp}(\rm aVDZ) \times \frac{E^{(2)}_{\rm disp}(\rm aVTZ)}{E^{(2)}_{\rm disp}(\rm aVDZ)} .
\label{exch-disp-gs}
\end{align}
where aVXZ = aug-cc-pVXZ, and it is assumed that the convergence of second-order polarization and exchange components with the basis set size is identical.

All presented SAPT(CAS) results include $\delta_{\rm HF}$ and $\delta_{\rm CAS}$ corrections for ground- and excited-state complexes, respectively, as well as scaled second-order exchange components defined in Eqs.~\eqref{exch-ind-gs}--\eqref{exch-disp-gs}. 
CAS+DISP is a sum of supermolecular CASSCF interaction energy and the dispersion energy, DISP=$E_{\rm disp}^{(2)}+E_{\rm exch-disp}^{(2)}$, computed in the same fashion as in SAPT(CAS), i.e., using the newly developed expression given in Eq.~\eqref{edisp} and the scaling relation from Eq.~\eqref{exch-disp-gs}.

\section{Results}

In Table~\ref{tab:SAPT} we present SAPT interaction energy decomposition for ground- and excited-state complexes. Regardless of the electronic state of the dimer, all systems can be classified as dispersion-dominated, with the $E^{(2)}_{\rm disp}/E^{(1)}_{\rm elst}$ ratio ranging from $2.8$ to $3.8$.

\begin{table*}
\centering
\caption{Components of the SAPT(CAS) interaction energy, including the $\delta_{\rm HF/CAS}$ correction, for ground- and excited-state complexes. The last column is a sum of all components. Differences of SAPT(CAS) energies between excited (ES) and ground states (GS), $ \Delta E_{\rm x} = E_{\rm x}(\textrm{ES}) - E_{\rm x}(\textrm{GS})$, are shown in the lower part of the Table. Energy unit is kcal$\cdot$mol$^{-1}$.}
\small
\begin{ruledtabular}
\begin{tabular}{l d{2.2} d{2.2} d{2.2} d{2.2} d{2.2} d{2.2} d{2.2} d{2.2}}
& \multicolumn{1}{c}{$E^{(1)}_{\rm elst}$} & \multicolumn{1}{c}{$E^{(1)}_{\rm exch}$} & \multicolumn{1}{c}{$E^{(2)}_{\rm ind}$} & \multicolumn{1}{c}{$E^{(2)}_{\rm exch-ind}$} & \multicolumn{1}{c}{$E^{(2)}_{\rm disp}$} & \multicolumn{1}{c}{$E^{(2)}_{\rm exch-disp}$} & \multicolumn{1}{c}{$\delta_{\rm HF/CAS}$} & \multicolumn{1}{c}{SAPT} \\ \hline
                     &	\multicolumn{8}{c}{ground state} \\
Benzene-Cyclopentane & -2.05 & 5.33 & -1.45 & 1.28 & -7.13 & 0.87 & -0.48 & -3.63 \\
Benzene-Neopentane   & -1.58 & 4.08 & -1.00 & 0.83 & -5.58 & 0.65 & -0.35 & -2.95 \\
AcOH-Pentane         & -1.54 & 4.20 & -1.05 & 0.83 & -5.55 & 0.57 & -0.28 & -2.82 \\
AcNH$_2$-Pentane     & -2.09 & 5.24 & -1.57 & 1.01 & -6.51 & 0.71 & -0.39 & -3.61 \\
Peptide-Pentane	    & -2.31 &  6.07 & -1.57 & 1.15 & -8.03 & 0.85 & -0.43 & -4.26 \\
                     &      &       &       &       &       &    &       &    \\
                     &	\multicolumn{8}{c}{excited state} \\
Benzene-Cyclopentane & -1.92 & 5.15 & -1.42 & 1.30 & -6.89 & 0.82 & -0.47 & -3.42 \\
Benzene-Neopentane   & -1.43 & 3.87 & -0.97 & 0.85 & -5.38 & 0.61 & -0.34 & -2.80 \\
AcOH-Pentane         & -1.53 & 4.12 & -1.03 & 0.91 & -5.61 & 0.56 & -0.27 & -2.85 \\
AcNH$_2$-Pentane     & -2.19 & 5.69 & -2.05 & 2.01 & -6.66 & 0.81 & -0.52 & -2.90 \\
Peptide-Pentane	     & -2.27 & 6.11 & -1.54 & 1.36 & -8.10 & 0.88 & -0.42 & -3.97 \\
 &	 &	&  &  &	 &	 &  & \\
       &	\multicolumn{8}{c}{excited state$-$ground state} \\
 & \multicolumn{1}{c}{$\Delta E^{(1)}_{\rm elst}$} & \multicolumn{1}{c}{$\Delta E^{(1)}_{\rm exch}$} & \multicolumn{1}{c}{$\Delta E^{(2)}_{\rm ind}$} &	\multicolumn{1}{c}{$\Delta E^{(2)}_{\rm exch-ind}$} & \multicolumn{1}{c}{$\Delta E^{(2)}_{\rm disp}$} & \multicolumn{1}{c}{$\Delta E^{(2)}_{\rm exch-disp}$} &  \multicolumn{1}{c}{$\Delta\delta_{\rm HF/CAS}$} & \multicolumn{1}{c}{SAPT} \\ \hline
Benzene-Cyclopentane	&  0.14 & -0.18 &  0.03 & 0.03  &  0.24 & -0.05 &  0.01 &  0.22 \\
Benzene-Neopentane	&  0.15 & -0.21 &  0.03 & 0.02	&  0.20 & -0.05 &  0.01 &  0.15 \\
AcOH-Pentane	        &  0.01 & -0.08 &  0.02 & 0.08	& -0.06 & -0.01 &  0.02 & -0.02 \\
AcNH$_2$-Pentane	& -0.09 &  0.45 & -0.48 & 1.00	& -0.15 &  0.10 & -0.12 &  0.71 \\
Peptide-Pentane	&  0.04 &  0.04 &  0.03 & 0.21	& -0.07 &  0.03 &  0.01 &  0.29 \\
\end{tabular}
\end{ruledtabular}
\label{tab:SAPT}
\end{table*}

As can be deduced from Table~\ref{tab:SAPT}, the most significant change in dispersion interactions upon transition from the ground to the excited state occurs in complexes of benzene (benzene$\cdots$cyclopentane and benzene$\cdots$neopentane). The effect amounts to $\Delta E^{(2)}_{\rm disp} \approx 0.2$~kcal/mol which corresponds to a decrease in the dispersion energy in the excited state. In both complexes, the redistribution of the electron density upon $\pi\to\pi^{*}$ excitation on benzene is accompanied by a non-negligible drop in the electrostatic attraction. The latter energetic effect is, however, cancelled by the simultaneous depletion of the exchange repulsion. Thus, decline of the dispersion energy contributes in a major way to the weakened net attraction in the excited state. We observed the same trends in dimers of benzene with \ce{H2O}, \ce{MeOH}, and \ce{MeNH2} studied in Ref.~\citenum{Jangrouei:22}.

Compared to benzene $\pi\to\pi^{*}$ systems, complexes of $n$-pentane involve a $n\to\pi^{*}$ exciton localized on the carbonyl group of the interacting partner (AcOH, \ce{AcNH2}, peptide). These systems exhibit an increase in the dispersion energy upon excitation which ranges from $-0.06$ to $-0.15$ kcal/mol (Table~\ref{tab:SAPT}). In AcOH$\cdots$pentane, the enhanced dispersion is comparable in magnitude with a concurrent decrease in the first-order Pauli repulsion, both of which contribute to the overall stabilization of the excited-state dimer. In contrast, in peptide$\cdots$pentane and \ce{AcNH2}$\cdots$pentane the net
repulsive components become stronger and outweigh the dispersion attraction, so that both complexes are more strongly bound in the ground state. For peptide$\cdots$pentane, the weakened interaction in the excited state can be attributed mainly to a significant increase of second-order exchange-induction ($\Delta E^{(2)}_{\rm exch-ind} = 0.21$~kcal/mol). The other interaction energy components undergo relatively minor changes, the only stabilizing effect of $-0.07$~kcal/mol is due to dispersion. In the case of \ce{AcNH2}$\cdots$pentane, increased static polarizability of acetamide in the excited state results in a stronger induction attraction (the net change in induction and $\delta$ corrections amounts to $-0.60$~kcal/mol). This, however, is countered by a steep rise in the repulsive components. In particular, exchange-induction and first-order exchange increase by $1.00$ and $0.45$~kcal/mol, respectively. Note that a similar pattern occurred in the methylamine$\cdots$peptide ($n-\pi^{*}$) interaction \cite{Jangrouei:22}.

\begin{figure*}
\centering
\includegraphics[width=0.8\textwidth]{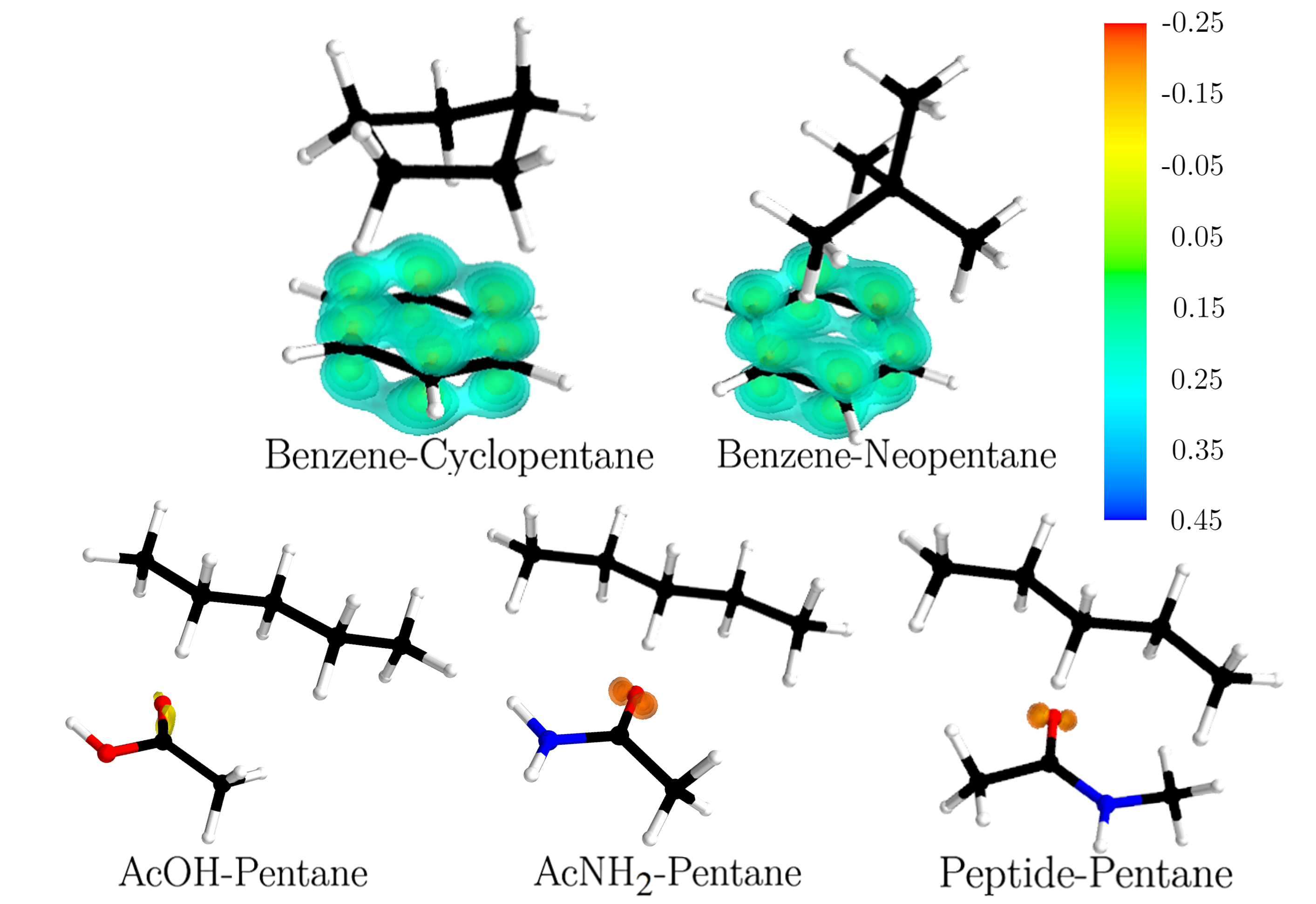}
\caption{Differences of the dispersion energy density $Q^{AB}\left(\mathbf{r}\right)$ between excited and ground states. Each isosurface consists of 6 contours of $Q^{AB}\left(\mathbf{r}\right)$ densities, generated to encompass $50\%$, $40\%$, $30\%$, $20\%$, $10\%$ and $1\%$ of the integrated $Q^{AB}\left(\mathbf{r}\right)$, respectively. Values on the color scale are reported in kcal$\cdot$mol$^{-1}$\AA$^{-3}$. Positive values correspond to regions where the dispersion energy density the in excited state is depleted (less negative) compared to the ground state.}
\label{fig:disp2}
\end{figure*}

Changes in the $E^{(2)}_{\rm disp}$ components are visualized in Figure~\ref{fig:disp2} using the difference between ground- and excited-state dispersion interaction density, $Q^{AB}(\mathbf{r})$, see Sec.~\ref{sec:vis}. Both the sign and magnitude of the effect are correctly captured---one observes a notable depletion of the dispersion density in $\pi-\pi^{*}$ complexes compared to a weaker accumulation for $n-\pi^{*}$ dimers. In agreement with the character of the underlying excitons, in the $\pi-\pi^{*}$ case majority of the $\Delta E^{(2)}_{\rm disp}$ term is delocalized over the benzene ring, while in $n-\pi^{*}$ dimers it is basically confined to the carbonyl group.

In Tables~\ref{tab:gs} and \ref{tab:es} we report total ground- and excited-state interaction energies, respectively, calculated at the CASSCF, CAS+DISP and SAPT(CAS) levels of theory. Addition of the dispersion energy to supermolecular CASSCF changes the character of the interaction from repulsive to attractive,  reducing mean errors by two orders of magnitude. In consequence, CAS+DISP results closely match the coupled-cluster (CC) reference \cite{Nakai:12} with mean absolute percentage errors (MA\%Es) of $0.8$\% for ground and $3.7$\% for excited-states. SAPT(CAS) performs similar to the CAS+DISP model (MA\%E values of $2.5$\% and $3.3$\% for ground and excited states, respectively). The DFT-based LRD model of Nakai et al. \cite{Nakai:12} combined with the LC-BOP functional \cite{Becke:88,Tsuneda:99,Iikura:01} is somewhat less accurate. The model underestimates interaction energies which amounts to mean errors at the level of 10\% (Tables~\ref{tab:gs}-\ref{tab:es}). Note, however, that DFT result were obtained in the 6-311++G(2d,2p) basis set.

Since the aug-cc-pVTZ basis set is not sufficient to saturate the dispersion energy with respect to the basis set size, the good agreement of both SAPT(CAS) and CAS+DISP with coupled-cluster is partially due to error cancellation (the CC values include CBS-extrapolated ground-state energies). Indeed, individual SAPT(CAS) energy components for ground-state complexes are systematically underestimated with respect to their SAPT(DFT) counterparts (see Tables~S2-S4 in the Supporting Information). This reflects the effective neglect of intramonomer electron correlation effects in the SAPT(CAS) approach \cite{Hapka:19a,Hapka:19b,Hapka:21}.

\begin{table*}
\centering
\caption{Ground state interaction energies in kcal$\cdot$mol$^{-1}$. 
CCSD(T)/CBS results from Ref.~\citenum{Rezac:13} are given as reference in the last column. The SAPT acronym refers to SAPT(CAS) results including the $\delta_{\rm HF}$ correction. TD-LC-BOP+LRD results are taken from Ref.~\citenum{Nakai:12}. Mean absolute errors (MAE) and mean absolute percent errors (MA\%E) are computed with respect to the reference.}
\small
\begin{ruledtabular}
\begin{tabular}{l d{2.2} d{2.2} d{2.2} d{2.2} d{2.2} }
    & \multicolumn{1}{c}{CASSCF} & \multicolumn{1}{c}{CAS} & \multicolumn{1}{c}{SAPT} & \multicolumn{1}{c}{TD-LC-BOP} & \multicolumn{1}{c}{Ref.~\citenum{Rezac:13}} \\
    &	       &  \multicolumn{1}{c}{+DISP}  &       & \multicolumn{1}{c}{+LRD}       &  \\ \hline
Benzene-Cyclopentane & 2.75 & -3.51 & -3.63 & -3.04 & -3.51 \\
Benzene-Neopentane   & 2.06 & -2.86 & -2.95 & -2.72 & -2.85 \\
AcOH-Pentane         & 2.18 & -2.86 & -2.82 & -2.49 & -2.91 \\
AcNH$_2$-Pentane     & 2.32 & -3.48 & -3.61 & -2.96 & -3.53 \\
Peptide-Pentane      & 2.95 & -4.23 & -4.26 & -3.47 & -4.26 \\
&       &       &      &      &	   \\
MAE     &  5.86 & 0.03 & 0.08 &  0.48 & \multicolumn{1}{c}{-} \\
MA\%E	&  \multicolumn{1}{c}{172} & 0.8  & 2.5  & 13.4  & \multicolumn{1}{c}{-} \\
\end{tabular}
\end{ruledtabular}
\label{tab:gs}
\end{table*}

\begin{table*}
\centering
\caption{Interaction energies in kcal$\cdot$mol$^{-1}$  for $\pi-\pi$* 
(benzene complexes) and $n-\pi$* (pentane complexes) excited states. The SAPT acronym refers to SAPT(CAS) results including the $\delta_{\rm CAS}$ correction. The Est.\ EOM-CCSD(T) values from Ref.~\citenum{Nakai:12} are given as reference in the last column. TD-LC-BOP+LRD results are taken from Ref.~\citenum{Nakai:12}. Mean absolute errors (MAE) and mean absolute percentage errors (MA\%E)  are computed with respect to the reference.}
\small
\begin{ruledtabular}
\begin{tabular}{l d{2.2} d{2.2} d{2.2} d{2.2} d{2.2} }
& \multicolumn{1}{c}{CASSCF}	& \multicolumn{1}{c}{CAS} & \multicolumn{1}{c}{SAPT} & \multicolumn{1}{c}{TD-LC-BOP} & \multicolumn{1}{c}{Ref.~\citenum{Nakai:12}} \\
&	     	&	\multicolumn{1}{c}{+DISP} 		&      &   \multicolumn{1}{c}{+LRD}     &      \\ \hline
Benzene-Cyclopentane & 2.77	& -3.31	&	-3.42	& -3.11	& -3.45	\\
Benzene-Neopentane   & 2.06	& -2.71	&	-2.80	& -2.79	& -2.86	\\
AcOH-Pentane	     & 2.27     & -2.78         &       -2.85	& -2.67	& -3.03	\\
AcNH$_2$-Pentane     & 3.11	& -2.74	&	-2.90	& -2.48	& -2.76	\\
Peptide-Pentane      & 3.18	& -4.05	&	-3.97	& -3.52         & -4.07	\\
 & & & & &	\\
MAE     & 5.91 & 0.12 & 0.10 & 0.32 & \multicolumn{1}{c}{-} \\
MA\%E	& \multicolumn{1}{c}{184}  & 3.7  &	3.3	 & 9.6  & \multicolumn{1}{c}{-} \\
\end{tabular}
\end{ruledtabular}
\label{tab:es}
\end{table*}

\section{Conclusions}
We have presented an algorithm for second-order dispersion energy calculations with multiconfigurational wave functions that scales with the fifth power the system size. Until now, $m^5$ scaling in coupled dispersion energy computations could only be achieved with single-determinant description of the monomers \cite{Bukowski:05,Podeszwa:06,Podeszwa:12,Xie:22}. The prerequisite for $m^5$ scaling with a multiconfigurational reference is that the number of active orbitals in wave functions of the monomers is considerably smaller compared to the number of virtual orbitals. In practice, this condition is typically fulfilled in interaction energy calculations performed using augmented basis sets.

The algorithm relies on the Extended RPA solver to obtain density response functions of the monomers and employs Cholesky decomposition of two-electron integrals. The key step involves the coupling-constant expansion of the response matrix projected onto the space spanned by the Cholesky vectors. The expansion follows from partitioning of the monomer Hamiltonian into  the zeroth-order partially interacting group-product-function Hamiltonian and  the remainder term scaled by the coupling constant $\alpha$. Consecutive terms of the response matrix expansion at $\alpha=0$ are calculated based on recursive relations proposed in Ref.~\citenum{Drwal:22}. Our numerical experience shows that truncation through  the eighth order is sufficient to achieve accuracy at the level of few $\mu E_h$.

The cost of the induction energy can be reduced from $m^5$ to $m^4$
in an infinite-order approach without the expansion in $\alpha$. This avoids the (small) numerical error
related to the truncation scheme (see Supporting Information).

To visually represent  the change in dispersion forces
upon vertical excitations, we have introduced
a spatial descriptor based on the proposed expression for the dispersion energy.
The underlying partition of the dispersion energy expression may be cast as a generalization of the approach first developed by Parrish and Sherrill \cite{Parrish:14a} for single-reference wave functions.

The new $m^5$ dispersion energy algorithm was employed together with state-averaged CASSCF wave functions to examine interactions involving localized excitons of the $\pi-\pi^{*}$ and $n-\pi^{*}$ type. For representation of the interaction energy, multiconfigurational dispersion energy was complemented either with SAPT(MC) \cite{Hapka:21} energy components or with supermolecular CASSCF interaction energy, the latter known as the CAS+DISP \cite{Hapka:20b} method. The dimers included up to eleven heavy atoms (between 800 and 900 basis set functions using the aug-cc-pVTZ basis set) which exceeded the capabilities of our original, $m^6$-implementation \cite{Hapka:19a}.

In line with earlier investigations \cite{Jangrouei:22}, SAPT decomposition shows that even in low-lying excited states the dispersion energy may be the driving force behind the stability of the complex. Hence, both accurate and efficient algorithms adequate for dispersion computations with multiconfigurational wave functions are mandatory. Spatial mapping of the dispersion energy density helps to identify regions affected most by the exciton. Visualizing the remaining SAPT(MC) energy components could aid the interpretation of energetic effects that occur upon electron density rearrangement in excited states. Work along this line is in progress.

\begin{acknowledgments}
This research was funded in whole or in part by National Science Center, Poland under grants no.\ 2019/35/B/ST4/01310 and no.\ 2021/43/D/ST4/02762. For the purpose of Open Access, the author has applied a CC-BY public copyright licence to any Author Accepted Manuscript (AAM) version arising from this submission.
This research was also funded by the European Centre of Excellence in Exascale Computing TREX - Targeting Real Chemical Accuracy at the Exascale. The project has received funding from the European Union’s Horizon 2020 - Research and Innovation program - under grant agreement no.\ 952165. 
\end{acknowledgments}

\bibliography{main}

\end{document}